\documentclass[12pt,preprint]{aastex}
\usepackage{times}

\shorttitle{Surface Magnetism Effects in Time-Distance}
\shortauthors{Zhao \& Kosovichev}

\begin{document}
\title{Surface Magnetism Effects in Time-Distance Helioseismology}
\author{Junwei Zhao and Alexander G. Kosovichev}
\affil{W. W. Hansen Experimental Physics Laboratory, Stanford University,
Stanford, CA94305-4085}
\email{junwei@quake.stanford.edu}

\begin{abstract}

Recent observations of helioseismic holography revealed that
magnetic fields that are inclined relative to the line-of-sight
direction could cause systematic variations in measured acoustic
phase shifts (hereafter, ``inclined magnetic field effect''), and
that the presence of surface magnetic field may shift the phases
and impair the coherence of acoustic waves (known as ``showerglass
effect''), thus complicating the interpretation of acoustic wave
propagation time through the solar interior. In this paper, we examine
how these two observational effects affect time-distance helioseismology
measurements in magnetic regions. It is confirmed that the inclined
magnetic field could cause variations in time-distance measured acoustic
travel times inside sunspot penumbra as well, however, inversions of the
measured times for the wave propagation show that this effect only slightly
shifts the location of negative sound-speed variations near the solar
surface, but basically does not change the inverted deeper interior
structures. Further measurements by use of continuum intensitygrams
and line-depth data from the Michelson Doppler Imager (MDI) onboard
{\it Solar and Heliospheric Observatory (SOHO)} illustrate that the
inclined magnetic field does not cause any obvious systematic travel time
variations in these observations. Regarding to the showerglass effect,
we find that outgoing and ingoing travel time perturbations through sunspots
from our typical time-distance measurements are significantly smaller
than those reported from helioseismic holography, and also strongly
depend on the propagation depth indicating deep changes. In addition,
our second-skip cross-correlation experiments demonstrate that inside
sunspots, the half of the double-skip travel times are very similar to
the mean single-skip travel times, indicating that acoustic signals
observed inside sunspots do not introduce detectable phase shifts
after applying a proper phase-speed filtering. We finally conclude
that these surface magnetism effects do not cause considerable systematic
errors in time-distance helioseismology of active regions.

\end{abstract}


\section{Introduction}
\label{sec1}

Local helioseismology has become an important tool to study solar
interior structures and dynamics. Significant progress has been
made recently, e.g., deriving subsurface structures and flow fields
of sunspots \citep{kos00, zha01, sun02}, inferring large scale subsurface
flows \citep{hab02, zha04, kom04}, and imaging solar far side active
regions \citep{lin00, bra01}, by use of different local helioseismological
techniques, including time-distance helioseismology, ring-diagram
analysis, acoustic imaging and helioseismic holography. Meanwhile,
with the advancement of scientific investigations, specific cautions
are also taken on interpreting local helioseismology observations,
which include efforts to refine time-distance measurement \citep{giz04},
to improve accuracy of modeling and interpreting observations \citep{bir04,
hin05}, and to address some potential systematic effects in measurements,
such as ``inclined magnetic field effect'' \citep{sch05}, ``showerglass
effect'' \citep{lin05a,lin05b}, and ``masking effect'' caused by
acoustic power deficit in sunspots \citep{raj05}.

Both the inclined magnetic field effect and showerglass effect were first
observed by use of the helioseismic holography technique. By measuring
phase shifts obtained from the so-called ``local control correlation''
phase-sensitive holography \citep{lin05b} inside sunspot penumbra
when the sunspot was at different locations on the solar disk,
\citet{sch05} found that ingression acoustic phase shifts, which
correspond to the ingoing travel times in time-distance measurement,
vary with different viewing angles. They suggested that these phase
shifts might be due to the inclination of magnetic field lines
relative to the line-of-sight direction. They argued that the
inclined magnetic field might cause an elliptic photospheric motion,
resulting in variations of the observed magnetoacoustic wave with
the viewing angle with respect to the field direction. The other,
showerglass effect, has been introduced by \citet{lin04, lin05a, lin05b}.
The authors argued that the surface magnetic field may shift the phases
of acoustic waves, and that the phase shifts function as a sort of
acoustic showerglass that impairs the coherence of seismic waves and
degrades images of subsurface anomalies. Unlike the inclined magnetic
field effect, which is caused by the inclination of magnetic field and
most noticeable in sunspot penumbra close to the solar limb,
the showerglass effect exists in wherever magnetic field is present
at solar surface.

Although these observational effects were found by the helioseismic holography
technique, we are interested to know whether the similar effects exist in
time-distance measurements, and how they affect the inversion results
from time-distance helioseismology. In this paper, we measure
the inclined magnetic field effect by the use of the time-distance technique
on different types of observations of solar oscillations, including
MDI Dopplergrams, intensitygrams and line-depth data \citep{sch95}, and
also perform inversions for one selected active region to examine how
this effect may affect inverted interior structures. These results are
presented in Section~\ref{sec2}. We then present in Section~\ref{sec3}
the showerglass effect measurements, and in Section~\ref{sec4} the
second-skip travel times measured from annulus-annulus cross-correlations,
which demonstrate that oscillation signals inside sunspots do not
introduce significant extra phase shifts to time-distance measurements.
Discussions and conclusions follow in Section~\ref{sec4} and \ref{sec5}.
However, it should be borne in mind that the motivation of this paper is not
to compare directly measurements from time-distance and helioseismic
holography techniques, but to assess how some measurement effects that
were reported from holography studies affect our typical time-distance
measurements.

\section{Inclined Magnetic Field Effect}
\label{sec2}

\subsection{Measurements and Inversions from Dopplergrams}
\label{sec2p1}

To study the inclined magnetic field effect, we select the same
sunspot in active region AR9026 as studied by \citet{sch05} during the
similar observational periods. This sunspot appeared from the solar
east limb on June 2, 2000 at the latitude of $19\fdg5$ in the Northern
hemisphere. Our time-distance analysis is different from
the holography analysis \citep{sch05} in several aspects, including
observational duration, annulus radius and width (or pupil size),
frequency bands selection, and the use of phase-speed filtering.
Time-distance helioseismology employs phase-speed filtering procedure
\citep[e.g.,][]{duv96}, which is designed to keep
all the acoustic waves that have the same first bounce distances and
similar wave propagation speed (therefore, traveling similar distances
with similar times), and filter out all other acoustic waves. Applying
phase-speed filtering makes the time-distance analysis with short
annulus radius and short annulus width possible, and significantly
improves the signal to noise ratio of time-distance measurement.
We selected three dates for analysis, and for each date, only
observations of the first 512 minutes of the day are used. The
$p$-modes frequency range covers approximately $2.5 - 5.5$ mHz
after filtering, though the acoustic power peaks at approximately $3.5$ mHz.
The acoustic travel times are measured between a central point and
the surrounding annulus, with the annulus size range of $3.7 - 8.7$ Mm.

\begin{figure}[!ht]
\epsscale{0.8} \plotone{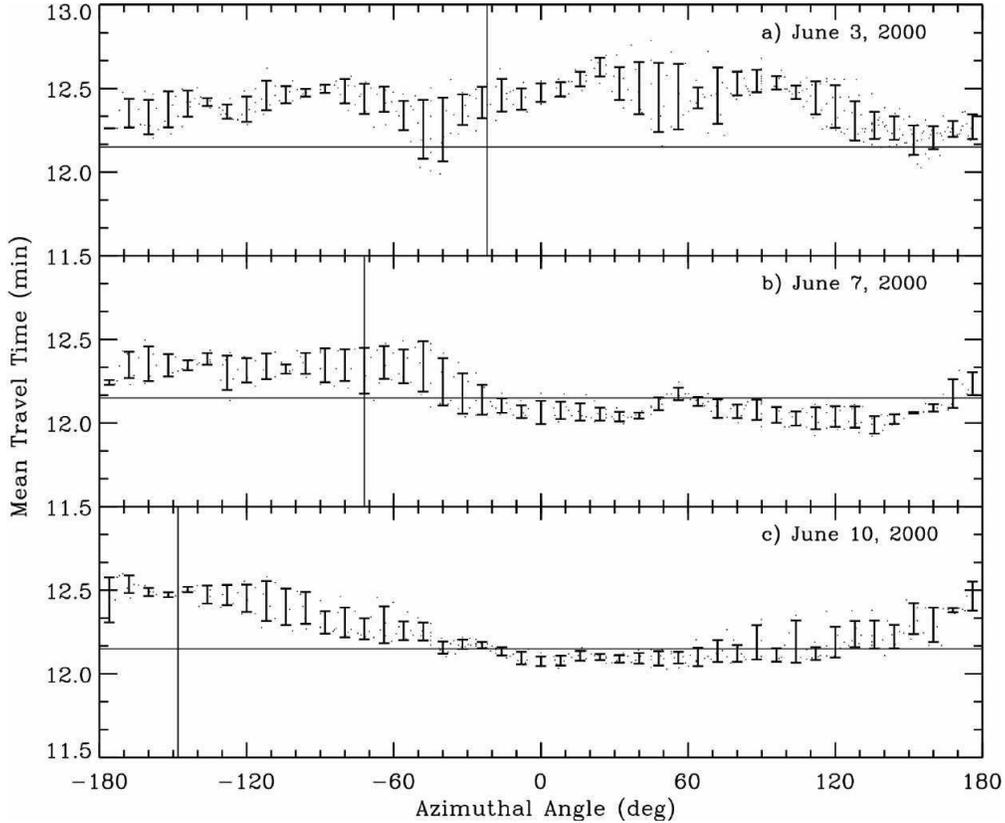} \caption{Mean acoustic travel times
in the penumbra of sunspot in AR9026, plotted as functions of the
azimuthal angle for three different observation dates. All pixels
inside the sunspot penumbra are used, and plotted as dots. For every
$4\degr$ interval in the azimuthal angle, the average and standard
deviation are computed and plotted as error bars. This sunspot is
located at latitude of $19\fdg5$N, and the vertical lines in (a),
(b) and (c) indicate the azimuthal angle of the solar disk center
relative to the center of the sunspot. The heliocentric distances
from the sunspot to the disk center are approximately $62\fdg0$,
$20\fdg5$, and $36\fdg8$ for (a), (b) and (c), respectively. The
horizontal line at the mean travel time of 12.16 minute is an
average acoustic travel time obtained from a quiet Sun region.}
\label{fg1}
\end{figure}

Following the data analysis procedure described by \citet{zha01}, we
compute the acoustic travel times inside the sunspot penumbra for
the selected observation periods.  Figure~\ref{fg1} presents the
mean travel times, which are averages of the outgoing and ingoing
travel times, as functions of the penumbral azimuthal angle for
different dates. The azimuthal angle starts from $0\degr$ in the West
of the sunspot umbra center, and increases counter-clockwise inside
the sunspot penumbra. Basically, the mean acoustic travel times
are longer inside the sunspot penumbra than these in the quiet Sun for this
specific annulus range that we use. It can be found that for each date,
the mean travel times vary with the azimuthal angle, with a magnitude
of approximately 0.2 minutes relative to the average value inside penumbra.
Additionally, the angular dependence of the variations of the mean travel
times is different for different dates, which implies that such
variations are not caused by the properties of the sunspot itself,
but by some systematic effects in the helioseismic measurements. More
examinations of different sunspots confirmed that such variations are
systematic measurement effects that are related to the projection effect,
rather than caused by the real variations in solar structures or
dynamics. Comparing Figure~\ref{fg1}
with Figure~2 in \citet{sch05}, one can tell that the travel time
variation trends are similar for the June 7 and 10 measurements, except that
the variation magnitude in our measurement is approximately 12 sec,
substantially smaller than that from the holography measurement, which
is of the order of 30 sec. Perhaps, the use of different measurement
annulus (or pupil) sizes, phase-speed filtering and different acoustic
frequencies in these techniques may have caused such differences.

\begin{figure}[!t]
\epsscale{1.0} \plotone{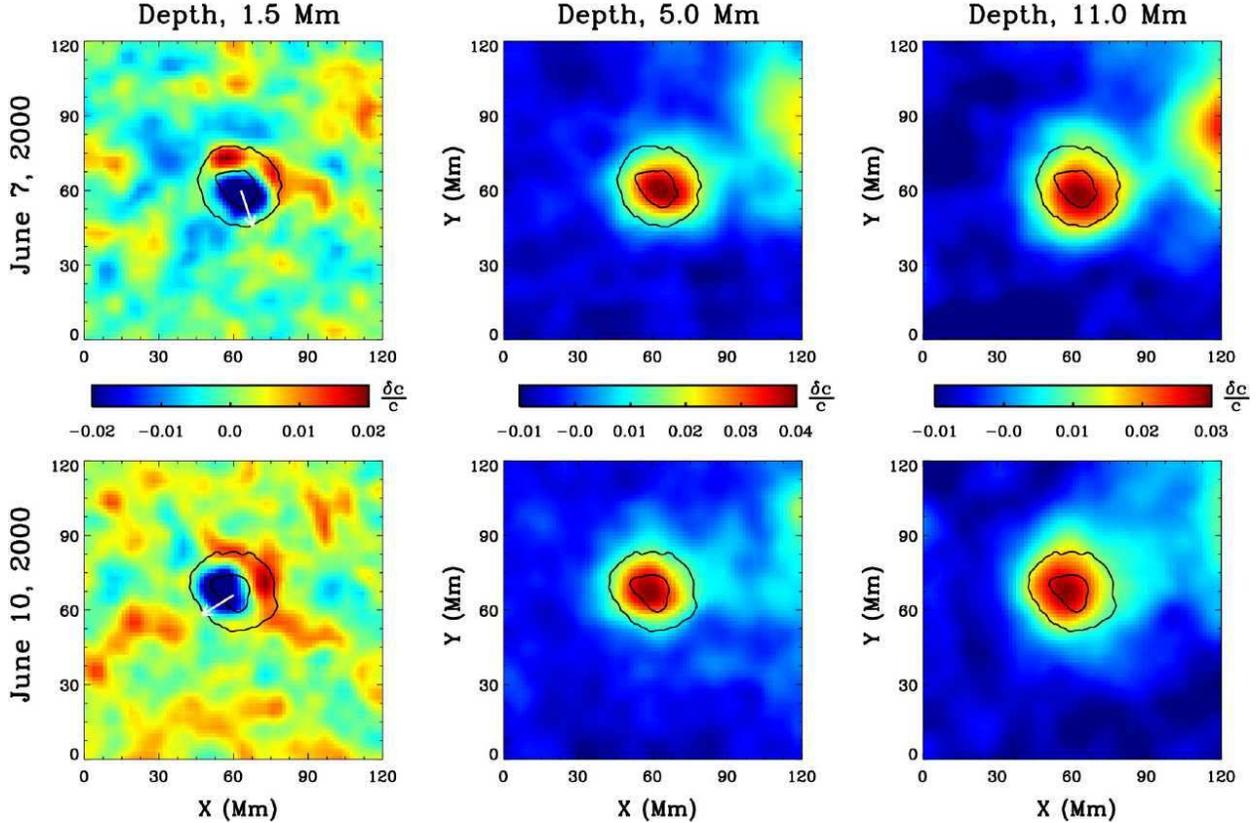} \caption{Sound-speed variations
inferred from time-distance inversions, shown at selected depths:
1.5 Mm in the left column, 5.0 Mm in the middle column, and 11.0 Mm
in the right column, for two dates: June 7, 2000 in the upper panel
and June 10 in the lower panel. Contours indicate the boundaries of
the sunspot umbra and penumbra, which are determined from MDI
continuum intensity observations. White arrows at the depth of 1.5
Mm on both dates point to the solar disk center. For different
dates, the image display color index is the same for the same depth,
with the color bars shown in the middle row. } \label{fg2}
\end{figure}

It is interesting to investigate how this measurement effect affects
our inversion results for sound-speed structures beneath the sunspot
\citep{kos00}. Since the inversion procedure involves integration over
large areas of different annuli and over a large depth range, as well
as averaging and smoothing, it is not obvious how much this systematic
measurement effect would change the inversion results.
Following the inversion procedures described in \citet{kos00} and
\citet{zha03}, we derive the subsurface sound-speed variation structures of
this sunspot from the travel time measurements for a total of 11
annuli, including the ones shown in Figure~\ref{fg1}.
Inversions for the June 3, 2000 observation are not performed because the
sunspot was located too close to the solar east limb.

The typical sound-speed variation structures \citep[e.g.,][]{kos00, zha03,
cou04, hug05} can be seen in Figure~\ref{fg2}, with negative sound-speed
variations close to the solar surface and positive sound-speed variations
below 5.0 Mm or so. Despite the evolutionary sunspot structure changes
that might have occurred from June 7 to June 10, some systematic changes
are still visible. It can be found that close to the surface, at the
depth of 1.5 Mm, the locations of negative sound-speed variations do not
coincide with the center of sunspot umbra, but slightly off from the
umbra towards the direction of the solar disk center; and additionally,
some positive sound-speed variations are found on the other side of
the umbra opposite to the disk center. Such changes are presumably
caused by the systematic inclined magnetic field effect.
However, at the depth of 5.0 Mm when the sound-speed variations become
positive, locations of major variations are coincident with the
locations of sunspot, including both umbra and penumbra. Even deeper,
at the depth of 11.0 Mm, the structures of sound-speed variations
do not change much, but again, locations may slightly
be off the center and move towards the direction of solar disk center.
Overall, when the sunspot was located on either side of the disk center, and
the inclined angles of magnetic field lines were different, but the
inverted interior structures of the sunspot remain largely unchanged
except near the solar surface.

\subsection{Measurements from MDI Continuum Intensity and Line-Depth Data}
\label{sec2p2}

MDI sometimes has simultaneous Dopplergram and continuum intensitygram
observations, or simultaneous Dopplergram and line-depth observations,
both with one minute cadence \citep{sch95}. It is interesting to
investigate how the inclination of the magnetic field may affect the
time-distance results obtained from the intensitygram and line-depth
solar oscillation data.

\subsubsection{Measurements from MDI Continuum Intensitygrams}

We select a sunspot with simultaneous MDI high resolution Dopplergram
and continuum intensitygram observations. This sunspot was located inside
active region AR8243, and the selected 512-minute observation period
is from 17:00UT, June 18 to 01:31UT, June 19, 1998, when the active region
was passing right above the solar disk center, at a latitude of
approximately $18\fdg0$N.

It is well known that the acoustic power computed from solar
intensitygrams is less stronger than that computed from Dopplergrams,
thus time-distance measurements from intensitygrams are noisier
\citep[e.g.,][]{sek01}. However, although the travel time maps computed from
the intensitygrams are noisier for short distances (small annuli), the
signatures of active regions and supergranulations are still clear,
and such measurements are sufficiently good to perform the analysis
similar to Section~\ref{sec2p1} in order to evaluate travel time
variations with the azimuthal angle inside the sunspot penumbra.

\begin{figure}[!t]
\epsscale{0.8} \plotone{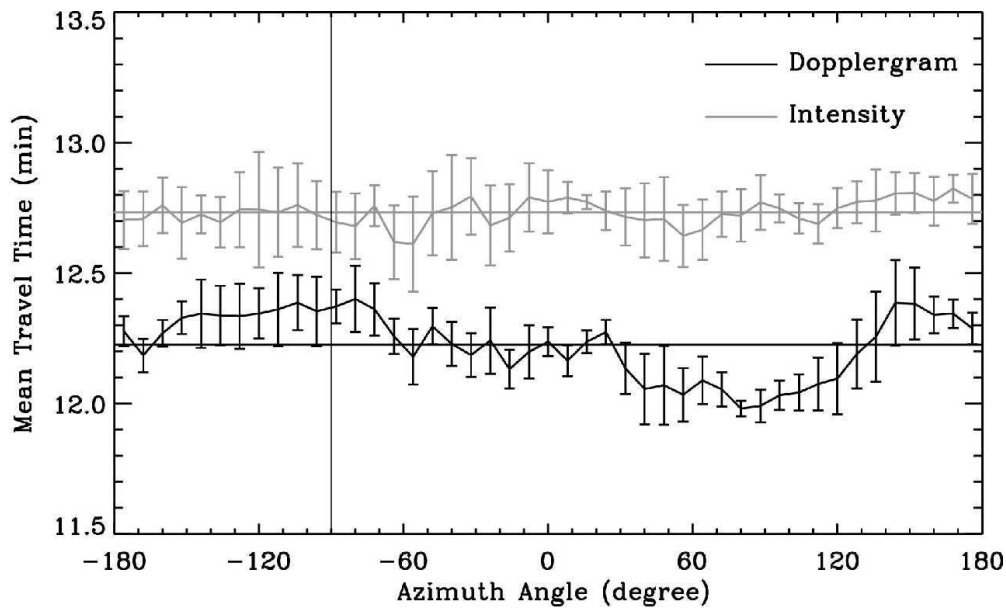} \caption{Mean acoustic travel time
variations inside the sunspot penumbra with the azimuthal angle,
measured from MDI Dopplergrams and intensitygrams for a sunspot
inside AR8243 of June 18-19, 1998. The black curve represents
results computed from the Dopplergrams and the light gray curve
represents results from the continuum intensitygrams. The error bars
are standard deviations as in Figure~\ref{fg1}. The dark and gray
horizontal lines indicate the averaged mean travel time inside the
penumbra from the Dopplergrams and intensitygrams, respectively. The
vertical line indicates the azimuthal angle of the solar disk center
relative to the center of the sunspot.} \label{fg3}
\end{figure}

Figure~\ref{fg3} shows the measurement results. The averages of the mean travel
times inside the sunspot penumbra measured from the two datasets are different,
so that the travel time measured from the continuum intensitygrams about
0.5 minutes longer than that the corresponding travel times from the
Dopplergrams. This is probably caused by the different acoustic power
distributions in the $k-\omega$ diagrams of these two kinds of observations,
which is well known in helioseismology. For example, the ratios of
acoustic powers in higher frequency and lower frequency are different,
and power spectra from Dopplergrams and intensitygrams display different
line asymmetries \citep{duv93}. It is still not very clear what causes
these differences, but it was suggested that these might be related to
the differences in the correlated components of the background noises
\citep{nig98}. The techniques to correct such travel time offset is
currently under investigation, and the main idea is to manipulate the
power spectra from intensitygrams to match the spectra from Dopplergrams
before computing acoustic travel times from the time-distance technique.
Nevertheless, Figure~\ref{fg3} shows that the acoustic travel times
measured from Dopplergrams have a cyclic 0.2 minutes azimuthal variations,
while the travel times from intensitygrams are basically invariant with
the azimuthal angle, and that the small variation of about 0.05 minutes
are likely related to the real structure variations or intensitygram
noises, rather than the systematic effects caused by the inclined
magnetic field.

\subsubsection{Measurements from MDI Line-Depth Data}

We select another sunspot with simultaneous MDI full-disk resolution
Dopplergram and line-depth observations. This sunspot was located
inside active region AR7973, and the selected 512-minute analysis
period was from 12:00UT to 20:31UT, June 25, 1996, when the sunspot was
approximately $5\fdg0$ past the central meridian and $7\fdg5$ above
the solar equator.

\begin{figure}[!t]
\epsscale{0.8} \plotone{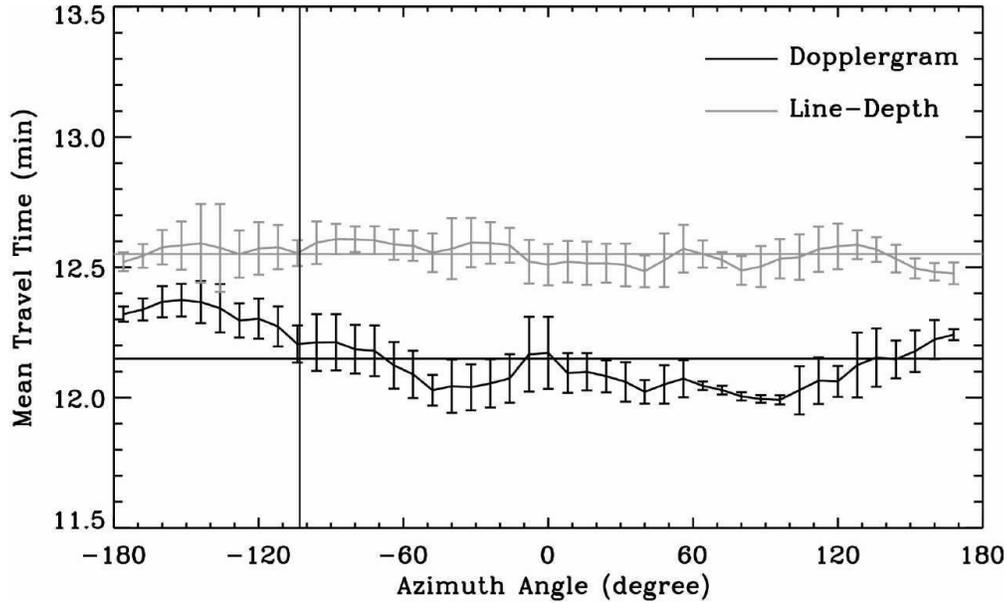} \caption{Same as Figure~\ref{fg3},
except that results are for a sunspot of AR7973 (June 25, 1996)
observed with the MDI full-disk resolution, and that the gray line
and error bars indicate results from the MDI line-depth data.}
\label{fg4}
\end{figure}

Similar to the results from intensitygrams, the acoustic power computed
from the line-depth data is also weaker than that computed from
Dopplergrams, and the time-distance results are also noisier. But
it is also true that the line-depth data give a stronger acoustic power
signal than the intensitygrams, and also give less noisier time-distance results
compared to the intensitygrams. The time-distance measurements performed
over such data are usable to evaluate travel time variations inside sunspot
penumbra. Figure~\ref{fg4} shows the results. The average mean travel time
measured from the MDI line-depth data are approximately 0.4 minutes
longer than that measured from the Dopplergrams, due to the similar reasons as
for the intensitygrams. Again, the acoustic travel times measured from the
Dopplergrams of this sunspot show 0.2 minutes variations with the azimuth,
but the travel times measured from the line-depth data do not have such
significant systematic variations.

\section{Showerglass Effect Measurements}
\label{sec3}

The showerglass effect introduced in helioseismic holography measurements is
exhibited as incoherent phase shifts of acoustic waves in active
regions, likely due to the presence of the surface magnetic field. One major
characteristic of this effect is that the amount of phase shifts, measured
from the so-called ``local ingression and egression control correlations'',
vary with magnetic field strength, and in particular, the local
ingression and egression control correlations exhibit asymmetric phase
shifts when magnetic field strength is strong
\citep[for details, see][]{lin05a}. It is quite clear that the local
control experiments used in helioseismic holography measurements are
a similar technique to measuring acoustic travel times by time-distance
helioseismology surface focusing scheme. Therefore, it is interesting
to carry out a similar analysis by use of time-distance technique, and
investigate how the presence of magnetic field may affect time-distance
measurements in active regions.

\begin{figure}[!ht]
\epsscale{1.0} \plotone{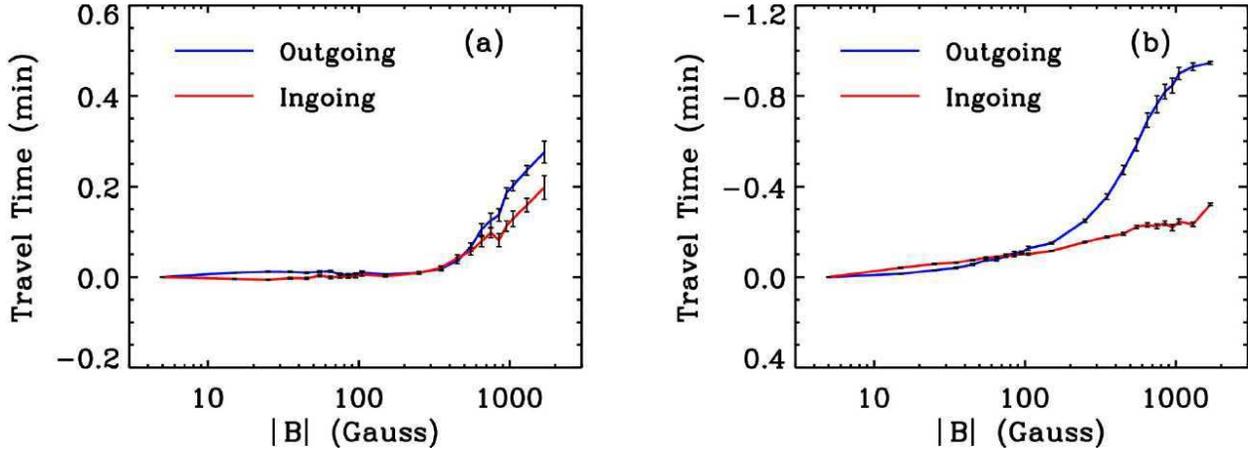} \caption{Outgoing and ingoing
travel times (relative to the quiet Sun) as functions of magnetic
field strength. The left panel shows results obtained with an
annulus range of $3.7 - 8.7$ Mm, and the right panel shows
measurements with an annulus range of $14.5 - 19.5$ Mm. Note the
different vertical scale in the two panels, and especially, it is
negative upward in the right panel. } \label{fg5}
\end{figure}

Figure~\ref{fg5} shows our measurements from AR 8243, the same Dopplergram
data as doing Dopplergram and intensitygram comparison analysis in
Section~\ref{sec2p2}, by use of two different annulus ranges. A significant
difference between our time-distance measurements and helioseismic holography
measurements \citep{lin05a} is that phase-speed filtering is applied in
our time-distance analysis while holography uses only frequency
filtering. Also, very short annulus widths (usually a few megameters)
are used in the time-distance analysis while holography often uses very large
pupil widths (approximately $30$ Mm). Comparing Figure~\ref{fg5} with
Figure~7 in \citet{lin05a}, one can find that the major similarities are
that the travel times (or phase shifts) vary with the magnetic field strength,
and that the outgoing and ingoing acoustic waves exhibit asymmetric travel
times when the magnetic field strength is large, for both short and long
annuli. Especially, the outgoing and ingoing travel times are very
asymmetric in Figure~\ref{fg5}b. However, the largest travel time
differences relative to the quiet Sun are approximately 0.3 minutes
in Figure~\ref{fg5}a and 1.0 minute in Figure~\ref{fg5}b, and these
numbers are significantly different from the result in \citet{lin05a},
which is approximately $180\degr$ in phase shift, i.e., 1.7 minutes.
Once again, these differences may be caused by the different frequency
bands, filtering technique and very different sizes of annulus (or pupil).

It is very important to note that the magnitude and sign of the travel
time variations are different for different travel distances. If travel
time variations are caused by the surface magnetism, then one may
expect that these variations do not significantly change with travel
distances. Therefore, the very different behaviors of the acoustic
travel times for different measurement annuli indicate that a large
fraction of the observed travel time deviations from the mean travel
time in the quiet Sun is due to the interior structures and dynamics
of active regions. The observed deviations of the travel times
in sunspots compared to the travel times in the quiet regions are
explained as changes in the sound-speed structures of sunspots with
depths, and the asymmetry in the outgoing and ingoing travel times
is explained as the advection effect due to subphotospheric flows.
The arguments that all travel time deviations from 0 are caused
by the showerglass effect, and that all such deviations be removed
from acoustic signals are not justified. Such corrections
will substantially underestimate sound-speed variations and flow
velocities in the sunspot interior. How and how much the surface
magnetism affects the acoustic travel times are definitely worth
more studies, and that will likely rely on numerical modelings.

\section{Second-Skip Experiment}
\label{sec4}

One important observation that supports the existence of showerglass
effect is that the phase shift, measured from helioseismic holography
ingression and egression correlations (corresponding to double-skip
travel time measurements of time-distance helioseismology) when the
holography focus plane is at the solar surface and located inside a
sunspot, is different from the sum of phase shifts measured from the
local ingression and egression control correlations. Measurements using
a second-skip time-distance technique showed similar discrepancies
\citep{bra97}. These observations supported the argument that acoustic
signals observed inside sunspot may cause outgoing and ingoing travel
time asymmetries, or introduce uncertainties to time-distance measured
travel times.

\begin{figure}[!ht]
\epsscale{0.8} \plotone{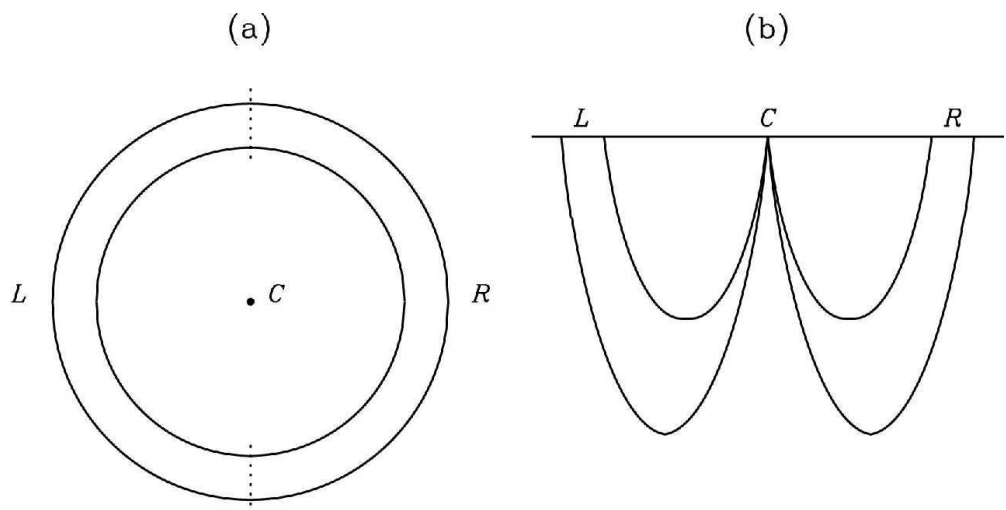} \caption{Above view (a) and side
view (b) of the time-distance measurement schemes. The traditional
single-skip acoustic travel times are measured by cross-correlating
signal from the central point {\it C} and the averaged signals
inside the whole annulus. The second-skip annulus-annulus
cross-correlation scheme, employed to derive the second-skip
acoustic travel times in this study, is to divide the whole annulus
into two semi-annuli, {\it L} and {\it R}, and then cross-correlate
the averaged signals inside {\it L} with those inside {\it R}. }
\label{fg6}
\end{figure}

\begin{figure}[!ht]
\epsscale{0.8} \plotone{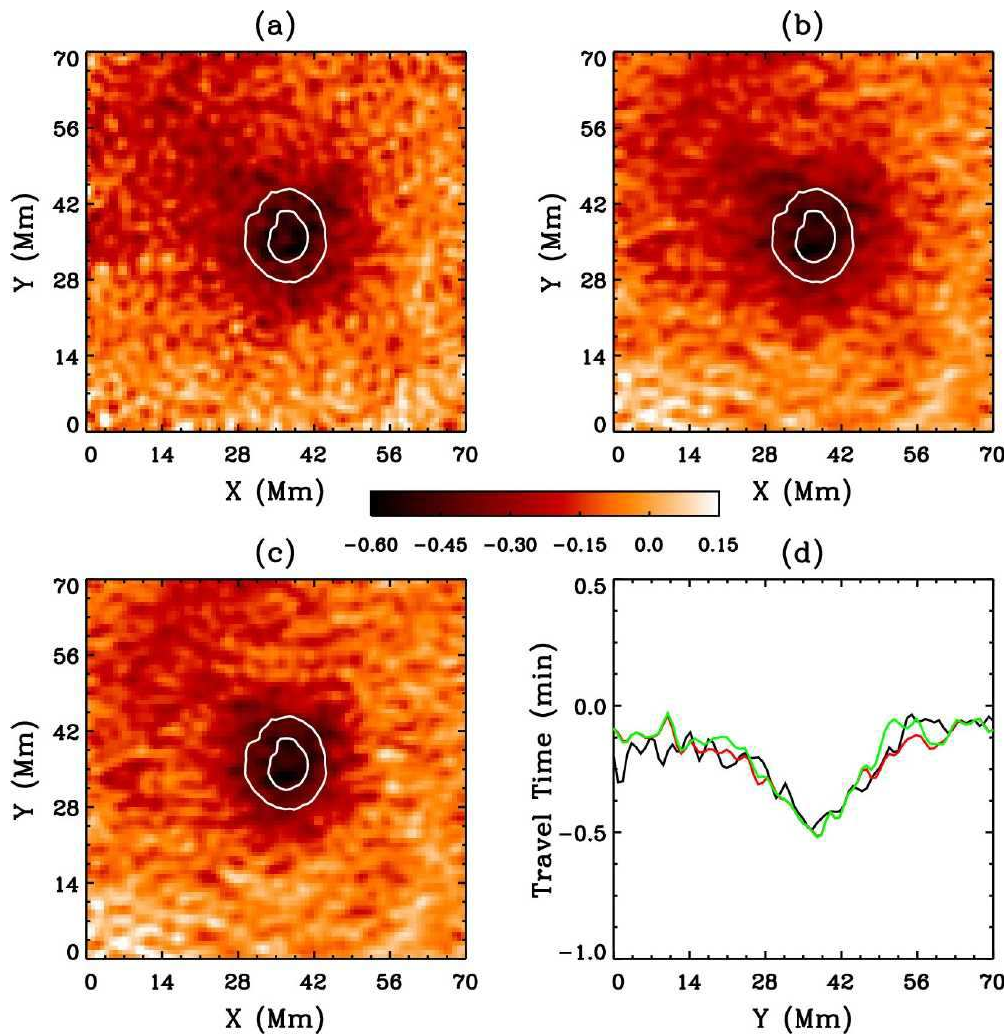} \caption{Comparison of the mean
travel time perturbation measurements using different measurement
schemes, including (a) traditional first-skip center-annulus
cross-correlations, (b) second-skip annulus-annulus
cross-correlations, and (c) second-skip annulus-annulus
cross-correlations when all the signals inside sunspot region are
not used in computations. Mean travel time perturbations are
relative to the quiet Sun, and in (b) and (c), travel times are
displayed after dividing 2. The annulus for making all three
measurements are the same, 32.2 Mm $-$ 41.3 Mm, and the color scales
for displaying are also the same, as shown in the color bar in the
middle. The unit for the color bar scale is minute. White contours
stand for the sunspot umbra and penumbra boundaries. To better
compare travel times, a vertical cut along X = 35 Mm of each image
is shown in (d), with the dark line from image (a), red line from
(b), and green line from (c). } \label{fg7}
\end{figure}

To avoid the use of pixels located inside the sunspot region as the central
point in traditional time-distance measurement (refer to Figure~\ref{fg6}),
\citet{hug05} used the second-skip time-distance measurements and performed
inversions, from which they obtained sound-speed perturbations beneath
sunspots similar to the results obtained from the traditional time-distance
measurements in the deep interior \citep{kos00, zha03}. A disagreement
between their second-skip analysis with the traditional one-skip analysis
near the surface may be due to the lack of near surface measurements when
performing second-skip measurements. However, their measurements employed
a scheme of center-double skip annulus cross-correlation (by
cross-correlating signals at the center with signals averaged from an
annulus around that center but with two acoustic skips), rather than the
scheme of annulus-annulus cross-correlation employed by \citet{bra97}, although
both schemes measure second-skip acoustic travel times.

\begin{figure}[!ht]
\epsscale{0.8} \plotone{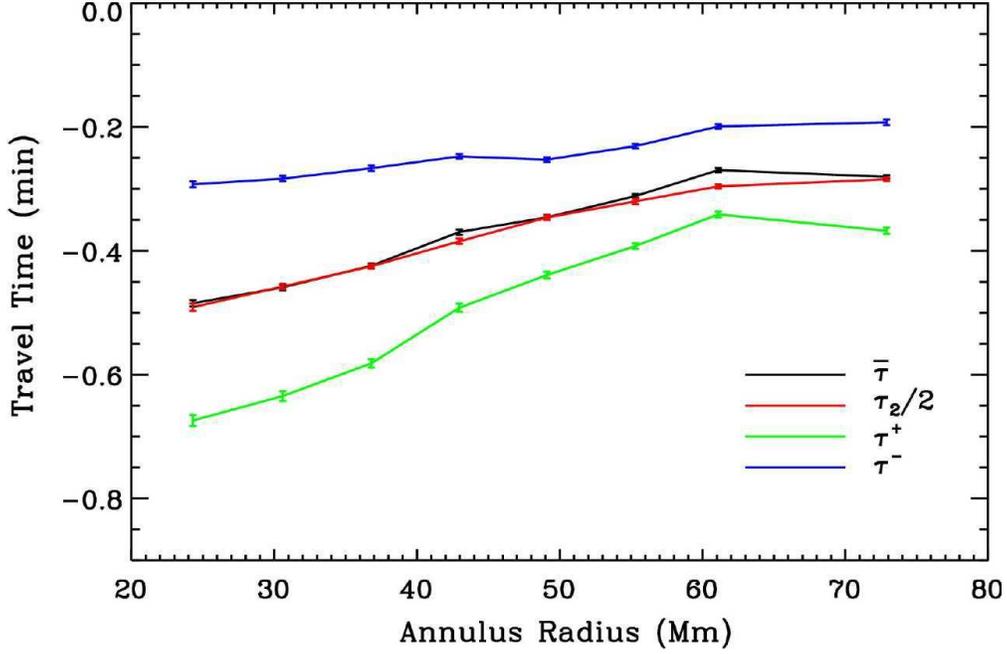} \caption{Averaged acoustic travel
times from the sunspot region shown as functions of annulus radius,
including outgoing travel times $\tau^{+}$ ingoing travel times
$\tau^{-}$, mean travel times $\overline{\tau} = (\tau^+ +
\tau^-)/2$, and half of double-skip travel times $\tau_2/2$. Again,
all the travel times are relative to the quiet Sun.} \label{fg8}
\end{figure}

We employ a similar scheme of annulus-annulus cross-correlation as used
by \citet{bra97} by dividing one circular annulus into two semi-annuli
and cross-correlating acoustic signals inside these two semi-annuli, as
illustrated in Figure~\ref{fg6}. After applying the phase-speed filtering,
commonly used in time-distance measurements, to the data of
AR8243 (same Dopplergram dataset as used in Section~\ref{sec2p2}), we
have measured the second-skip travel times for an annulus with the inner
radius of 32.2 Mm and the outer radius of 41.3 Mm, both of which are
larger than the distance from the sunspot umbra center to the boundary
of sunspot penumbra. It is worthwhile pointing out here, that the
oscillation power reduction (``masking'')
effect has been corrected based on the acoustic power map of the studied
region \citep{raj05} before measuring the acoustic travel times.
In addition to this second-skip experiment, another useful experiment
is to compute acoustic travel times through the annulus-annulus
cross-correlations, but without using oscillation signals inside the
entire sunspot region. This experiment is feasible because the mean
signals for computing cross-correlations are averaged from two semi-annuli,
which are not completely inside the sunspot, where data are not used,
for all cases that are computed in this paper. However, such experiment
cannot be done for the traditional center-annulus computations, because
pixels inside sunspots must be used as central points in these
computations. Figure~\ref{fg7} displays our travel time measurements
from the traditional single-skip center-annulus correlations, the
second-skip annulus-annulus correlations, and the
second-skip annulus-annulus correlations after zeroing
out signals inside the sunspot. It is quite clear that, from
Figure~\ref{fg7}a and b, these two different techniques give very
similar results in both measured structures and magnitude of travel times.
Figure~\ref{fg7}c once again confirms that with or without using
signals inside the sunspot does not change our acoustic travel time
measurements inside the sunspot, the region where our research interests are,
although travel times outside the active region are slightly enhanced.
This can be more clearly seen from Figure~\ref{fg7}d.

Selecting different annulus ranges, 19.4 Mm $-$ 29.3 Mm, 26.4 Mm $-$ 34.7 Mm,
32.2 Mm $-$ 41.3 Mm, 38.8 Mm $-$ 47.1 Mm, 44.2 Mm $-$ 54.1 Mm, 50.8 Mm $-$
59.9 Mm, 57.0 Mm $-$ 65.3 Mm and 67.7 Mm $-$ 76.8 Mm, outgoing
travel times $\tau^{+}$ and ingoing travel times $\tau^{-}$ from the
traditional single-skip, and travel times $\tau_{2}$ from the
second-skip annulus-annulus cross-correlations are calculated from
each annulus radius after applying the same phase-speed filtering.
Furthermore, mean travel times $\overline{\tau}$ are computed by averaging
the outgoing and ingoing travel times. The variations of these travel times
inside sunspot penumbra boundary as functions of annulus radii, defined
as the median of the annulus inner and outer radii, are presented in
Figure~\ref{fg8}. All travel times are plotted relative to the quiet
Sun acoustic travel times, which are measured for outgoing, ingoing,
mean and double-skip cases separately and are slightly different.
From Figure~\ref{fg8}, it is quite clear that inside sunspot region,
the ingoing travel times are often longer than the outgoing travel times.
More importantly, in contrast to the result that half of double-skip travel
times are closer to the ingoing travel times \citep{bra97}, it is
found in our measurements that the half of double-skip travel times
are very similar to the mean travel times, or almost same in some
annulus radii. This shows that the results of time-distance helioseismology
are not corrupted by using acoustic signals inside sunspots.

\section{Discussion}
\label{sec5}

In this paper, we have presented time-distance measurements of the
inclined magnetic field effect and the showerglass effect, both of which
were introduced and studied by use of the helioseismic holography
technique \citep{sch05, lin05a, lin05b}, and found these effects
are less significant in time-distance helioseismology, and do not
significantly affect the inversion results. By the second-skip
experiment, we have also found that acoustic signals inside sunspots carry
useful information and do not corrupt the time-distance results. Meanwhile,
it is worthwhile to point out here that, the purpose of this paper is
not to compare measurements from helioseismic holography and time-distance
helioseismology techniques, but to assess how some measurement effects, which
were reported in acoustic holography studies, affect the typical
time-distance measurements, and time-distance results that were reported
previously \citep[e.g.,][]{kos00,zha01}.

\subsection{Inclined Magnetic Field Effect}

Our time-distance measurements for the areas with magnetic field lines that are
inclined relative to the line-of-sight direction has confirmed the finding
by \citet{sch05}, that the presence of inclined magnetic field could
cause a variation of measured acoustic travel times by an amount of
approximately 0.2 minutes. It has been of significant interest to
investigate how this measurement effect might affect the inversions
of sunspot's interior structures. The inversion results for a selected
active region when this region was located at different positions on
the solar disk have revealed that, this observational effect may only slightly
shift the location of the negative sound-speed region, which is often
located close to the solar surface, towards the solar disk center.
The inversion results in the deeper interior, approximately below 5.0 Mm,
remain largely unaffected. Therefore, this measurement
effect does not change the major inversion result of the basic
sunspot structure that negative sound-speed perturbations are often
located in the shallow depth of the sunspot, and positive sound-speed
perturbations are beneath approximately 4 - 5 Mm.

On the other hand, time-distance measurements for MDI continuum
intensitygrams and line-depth data have shown that the mean travel times
inside sunspots' penumbra do not vary with the azimuthal angle. This
is quite an interesting result. It may indicate that the acoustic
information carried by Doppler velocity, which is a line-of-sight projection
of three-dimensional plasma velocity, is altered due to the interaction
between plasma motions and the transverse magnetic field, or due to
inaccuracies in the Doppler shift measurements in the presence of
inclined magnetic field. However, it seems that from our measurements,
acoustic information carried by
continuum intensitygrams and line-depth data, which are more related to
scalar physics quantities such as temperature and density, is not
altered by the presence of transverse magnetic field. This seems to
suggest that intensitygrams and line-depth data are better suited for
time-distance or other helioseismic holography studies, however, the
poorer signal-to-noise ratio of these data discourages a more general use.
The efforts for better understanding this observational effect, and
for designing correction techniques, most likely empirical, to correct
this effect in analysis of Dopplergram observations are ongoing.

\subsection{Showerglass Effect}

One major characteristic of the showerglass effect introduced in the
helioseismic holography technique is that the local ingression and egression
control measurements have larger phase shifts in magnetic regions,
and also exhibit asymmetric phase shifts \citep{lin05a, lin05b}.
Our time-distance measurements (Figure~\ref{fg5}) have
shown that, for the short annulus radius, acoustic waves inside
active regions have a longer travel time, an order of 0.3 minutes, than
the quiet Sun, for both the outgoing and ingoing travel times. While for
the longer annulus radius, the outgoing travel time inside active regions
can be up to 1.0 minute shorter than in the quiet region, and the ingoing
travel time is only about 0.3 minutes shorter. The asymmetry in outgoing
and ingoing travel times for both cases is obvious. However, all these
numbers are significantly smaller than the phase shifts measured by
the helioseismic holography technique \citep{lin05a}, though the differences
may come from different frequency bands and very different annulus
(pupil) sizes.

It was suggested that acoustic signals inside active regions were phase
shifted by the magnetic field \citep{lin05a}. However, it is demonstrated
from both Figures~\ref{fg5} and \ref{fg8}, that for different annulus radii,
the acoustic travel times inside active regions measured from time-distance
technique vary significantly, from $-1.0$ to $0.3$ minutes.
That is, phase shifts vary greatly depending on the annulus radii used to
make measurements, or in other words, phase shifts depend largely on
the depth, although one would expect that phase shift should
remain relatively constant or varies little if a significant amount of
phase shift is caused by the presence of surface magnetism rather
than by the sunspot interior structures and dynamics. Therefore,
it is reasonable to believe that a large fraction of travel time anomalies
measured by the time-distance method inside active regions are caused by
real subphotospheric structures and dynamics below sunspots rather than only
by the artifacts of sunspot surface magnetism. Certainly, the showerglass
effect itself, how it may affect various types of helioseismic measurements,
and how to account for this effect when it is significant are
worth of further studies, including observational, theoretical and
numerical studies.

\subsection{Second-Skip Experiment}

Our experiment of measuring the second-skip acoustic travel times by
annulus-annuls cross-correlations helps to prove that the observed
acoustic signals inside active regions provide useful information
and do not corrupt time-distance results. The mean travel times through
sunspots region, computed by averaging the outgoing and ingoing travel times,
are expected to be equal to half of the double-skip travel times, $\tau_2/2$,
because they actually measure the same signals along same ray paths
passing through the sunspot interior. Previous measurements by \citet{bra97}
found that $\tau_2/2$ was closer to the ingoing travel time
than to the mean travel time, which led him to question the use of
acoustic signals inside sunspots. This result was also considered as
a supportive evidence for the showerglass effect.
It should be pointed out here that the time-distance measurements by
\citet{bra97} did not apply commonly used filters that are designed to
filter out unuseful acoustic information (like the phase-speed filter
used in time-distance helioseismology), complicating interpretations of the
measurements. Nevertheless, our measurements (Figures~\ref{fg7} and
\ref{fg8}) have shown clear agreement of $\tau_2/2$ with mean travel
times $\overline{\tau}$. This obviously indicates that the use of
acoustic oscillation signals inside sunspots, as often done in traditional
time-distance measurements, do not introduce detectable acoustic
phase shifts that may corrupt the time-distance measurements and
inversion results. Probably, the differences of our measurements
and results of \citet{bra97} are caused by the use of phase-speed
filtering and different annulus radii and widths. Without the filtering,
mixed acoustic modes and different bounces of signals may come into the
measurements and complicate the analysis.

Therefore, the double-skip experiment performed in this paper suggests
that the time-distance measurements, after applying phase-speed filtering,
are basically self-consistent even inside highly magnetized sunspot regions.
It is also noteworthy that, by measuring acoustic signals through
the sunspot interior, \citet{duv95} already confirmed that sound-speed
in the sunspot deeper interior are faster, consistent with measurements
using observed oscillations inside sunspots. And also, by employing a
center-second skip annulus cross-correlation scheme, \citet{hug05} showed
that inversion results of deep sunspot interior structures from such
measurements agree with the previous time-distance inversion results.

\section{Conclusion}
\label{sec6}

In this paper, by use of the time-distance helioseismology technique,
we have examined two observational effects previously found by helioseismic
holography: the inclined magnetic field effect and showerglass effect
\citep{sch05, lin05a, lin05b}. We have confirmed the existence of the
inclined magnetic field effect in the time-distance analysis using
solar oscillation data from MDI Dopplergrams, but found that this
effect does not exist in the simultaneous continuum intensitygrams and
line-depth observations. Inversions of the time-distance measurements
of the MDI Doppler data reveal that inclined magnetic field effect is
practically insignificant for the helioseismic inferences of the basic
sunspot structure. The inverted sunspot structures just beneath the
surface may be slightly shifted when the sunspot is not located close
to the solar disk center, but deeper structures remain largely unaffected.
For the showerglass effect, we find that for different annulus radii,
time-distance measurements give different mean acoustic travel times inside
sunspot, which implies that what time-distance helioseismology measures
are largely the sunspot interior structure and dynamics rather than acoustic
signal phase shifts inside magnetized regions as suggested by \citet{lin05a}.
The second-skip time-distance experiment also convincingly demonstrates
that the double-skip acoustic travel times, measured without using
signals inside active regions, are in nice agreement with the mean
single-skip acoustic travel times obtained with using signals inside
active regions, which indicates that the solar oscillation signals
inside sunspots are useful for time-distance helioseismology measurements.
We believe that the numerical studies of the wave interactions with
the magnetic field will help us better understand local helioseismological
measurements.

\acknowledgments

We thank Drs.~Tom Duvall, Doug Braun and Charlie Lindsey for useful
discussions and comments on some parts of this research,
and we also thank them for thoroughly reading this manuscript and
providing valuable comments. We also thank an anonymous referee
for some constructive comments to improve the quality of this
paper. This research is partly supported by ``Living With a Star''
TR\&T program of NASA. The MDI/{\it SOHO} project is supported by
NASA grant NAG5-10483 to Stanford University. {\it SOHO} is a project
of international cooperation between ESA and NASA.


\begin{thebibliography}{}

\bibitem[Birch \& Felder(2004)]{bir04} Birch, A. C., \& Felder, G.
2004, \apj, 616, 1261

\bibitem[Braun(1997)]{bra97} Braun, D. C. 1997, \apj, 487, 447

\bibitem[Braun \& Lindsey(2001)]{bra01} Braun, D. C., \& Lindsey, C.
2001, \apjl, 560, L189

\bibitem[Couvidat et al.(2004)]{cou04} Couvidat, S., Birch, A. C.,
Kosovichev, A. G., \& Zhao, J. 2004, \apj, 607, 554

\bibitem[Duvall(1995)]{duv95} Duvall, T. L., Jr. 1995, in ASP Conf. Ser. 76,
Helio- and Asteroseismology, ed R. K. Ulrich, E. J. Rhodes, Jr., \&
W. D\"{a}ppen (San Francisco: ASP), 465

\bibitem[Duvall et al.(1996)]{duv96} Duvall, T. L., Jr., D'Silva,
S., Jefferies, S. M., Harvey, J. W., \& Schou, J. 1996, \nat,
379, 235

\bibitem[Duvall et al.(1993)]{duv93} Duvall, T. L., Jr., Jefferies, S. M.,
Harvey, J. W., Osaki, Y., \& Pomerantz, M. A. 1993, \apj, 410, 829

\bibitem[Gizon \& Birch(2004)]{giz04} Gizon, L., \& Birch, A. C.
2004, \apj, 614, 472

\bibitem[Haber et al.(2002)]{hab02} Haber, D. A., Hindman B. W.,
Toomre, J., Bogart, R. S., Larsen, R. M., \& Hill, F. 2002, \apj,
570, 855

\bibitem[Hindman et al.(2005)]{hin05} Hindman, B. W., Gough, D.,
Thompson, M. J., \& Toomre, J. 2005, \apj, 621, 512

\bibitem[Hughes, Rajaguru, \& Thompson(2005)]{hug05} Hughes, S. J.,
Rajaguru, S. P., \& Thompson, M. J. 2005, \apj, 627, 1040

\bibitem[Komm et al.(2004)]{kom04} Komm, R., Corbard, T., Durney, B. R.,
Gonz\'{a}lez-Hern\'{a}ndez, I., Hill, F., Howe, R., \& Toner, C.
2004, \apj, 605, 554

\bibitem[Kosovichev, Duvall, \& Scherrer(2000)]{kos00} Kosovichev, A. G.,
Duvall, T. L., Jr., \& Scherrer, P. H. 2000, \solphys, 192, 159

\bibitem[Lindsey \& Braun(2000)]{lin00} Lindsey, C., \& Braun, D. C.
2000, Science, 287, 1799

\bibitem[Lindsey \& Braun(2004)]{lin04} Lindsey, C., \& Braun, D. C.
2004, \apjs, 155, 209

\bibitem[Lindsey \& Braun(2005a)]{lin05a} Lindsey, C., \& Braun, D. C.
2005a, \apj, 620, 1107

\bibitem[Lindsey \& Braun(2005b)]{lin05b} Lindsey, C., \& Braun, D. C.
2005b, \apj, 620, 1118

\bibitem[Nigam et al.(1998)]{nig98} Nigam, R., Kosovichev, A. G.,
Scherrer, P. H., \& Schou, J. 1998, \apjl, 495, L115

\bibitem[Rajaguru, Zhao, \& Duvall(2005)]{raj05} Rajaguru, P., Zhao, J.,
\& Duvall, T. L., Jr. 2005, AGU Joint Assembly, SP11B-05

\bibitem[Scherrer et al.(1995)]{sch95} Scherrer, P. H., et al. 1995,
\solphys, 162, 129

\bibitem[Schunker et al.(2005)]{sch05} Schunker, H., Braun, D. C.,
Cally, P. S., \& Lindsey, C. 2005, \apjl, 621, L149

\bibitem[Sekii et al.(2001)]{sek01} Sekii, T., et al. 2001, In: Proc.
of the SOHO 10/GONG 2000 Workshop: Helio- and asteroseismology at the dawn
of the millennium, ed. A. Wilson (Noordwijk: ESA), 327

\bibitem[Sun et al.(2002)]{sun02} Sun, M. T., Chou, D. Y., \& the
TON Team 2002, \solphys, 209, 5

\bibitem[Zhao \& Kosovichev(2003)]{zha03} Zhao, J., \& Kosovichev, A. G.
2003, \apj, 591, 446

\bibitem[Zhao \& Kosovichev(2004)]{zha04} Zhao, J., \& Kosovichev, A. G.
2004, \apj, 603, 776

\bibitem[Zhao, Kosovichev, \& Duvall(2001)]{zha01} Zhao, J., Kosovichev,
A. G., \& Duvall, T. L., Jr. 2001, \apj, 557, 384

\end{thebibliography}
\end{document}